\documentclass{PoS}
\usepackage{graphicx}

\title{Compact source resolution and rapid variability in Arp220}

\ShortTitle{Compact source resolution and rapid variability in Arp220}

\author{\speaker{Fabien Batejat}$^a$, John E. Conway$^a$, Philip J. Diamond$^b$, Rodrigo Parra$^c$, Colin J. Lonsdale$^d$ and Carol J. Lonsdale$^e$\\
\llap{$^a$}Onsala Space Observatory, SE-43992 Onsala, Sweden\\
\llap{$^b$}CSIRO Astronomy and Space Science, Marsfield, NSW 2122, Australia\\
\llap{$^c$}CONICYT, Canad\'a 308, Santiago, 750-0788, Chile\\
\llap{$^d$}MIT Haystack Observatory, Westford, MA 01886, USA\\
\llap{$^e$}National Radio Astronomy Observatory, Charlottesville, VA 22903, USA\\
E-mail: \email{fabien.batejat@chalmers.se}}

\abstract{We present multi-epoch global VLBI observations at 2~cm, 3.6~cm and 6~cm of the compact radio sources in Arp220. We resolve many sources and estimate sizes, expansion velocities and source classes. We find most source properties are consistent with them being either radio supernovae (SNe) or supernova remnants (SNRs). We extend the luminosity-diameter relation for SNRs to very small sources and argue this supports models where shell magnetic fields are internally amplified. We also detect one probable SN/SNR transition object candidate and three highly variable sources with possible superluminal motion (approximately $4c$) of jet-like features near rapidly varying almost stationary components. These enigmatic sources, which show similarities to the recently discovered superluminal source in M82, might be associated with an AGN or a new radio source class (e.g. intermediate-mass black holes or micro-blazars).}

\FullConference{10th European VLBI Network Symposium and EVN Users Meeting: VLBI and the new generation of radio arrays \\
		 September 20-24, 2010\\
		 Manchester, UK}

\begin{document}

\section{Introduction} \label{sec:intro}
\subsection{Arp220} \label{sec:arp220}
Arp220 is the result of the collision of two galaxies now in the process of merging and has two distinct nuclei separated by approximately 350~pc~\cite{Norris88}. It is also the closest ULIRG (Ultra Luminous Infra-Red Galaxy) from the Earth, located 77~Mpc away. It has a far infrared luminosity of $\mathrm{L_{FIR}} \sim 1.3\times10^{12}\ \mathrm{L}_{\odot}$~\cite{Soifer87} which is more typical of star-forming galaxies at redshift $z = 1$ and has a similar star-formation density per unit area as redshift $z = 6$ proto-galaxies.
Smith et al 1998~\cite{Smith98} made the first detection at 18~cm of compact objects associated with the starburst. These objects were first detected at higher frequencies (13~cm, 6~cm and 3.6~cm) by Parra et al 2007~\cite{Parra07} who concluded that they are likely a mixture of SNe and SNRs.

\subsection{Supernovae and supernova remnants} \label{sec:SNe&SNRs}
In order to classify the sources observed in Arp220 it is important to understand the differences between SNe and SNRs and their radio emission mechanisms. The radio emission of a SN is the result of the interaction of the blast wave from the SN explosion with the ionised circumstellar medium (CSM). Relativistic particles in the SN blast wave interacting with a magnetic field produce non-thermal synchrotron emission which is absorbed by foreground thermal electrons. The amount of absorption decreases with time as the distance between the traveling shock and the edge of the CSM wind-blown bubble decreases. The radio light-curves characteristically peak from higher to lower frequencies followed by a power law decline at each frequency.

When the shock wave reaches the edge of the wind-blown bubble it starts interacting with the denser interstellar medium (ISM) gas. The SN becomes then a SNR. In a first phase the SNR expands at a nearly constant velocity radiating thermal X-rays and synchrotron emission. The light-curves increase at all frequencies because of the acceleration of relativistic electrons at the shock front created where the SNR meets the ISM. When the mass of ISM swept up by the shock wave is the same as the initial mass of the stellar ejecta the remnant enters the Sedov-Taylor phase and begins to decelerate. At this point it reaches its maximum radio luminosity. There exist different theories describing the radio emitting behavior of the SNR from this point on (see \S~\ref{sec:LDrel}).

\begin{figure} [ht]
\begin{center}
\includegraphics[width=0.55\textwidth, height=0.4\textwidth]{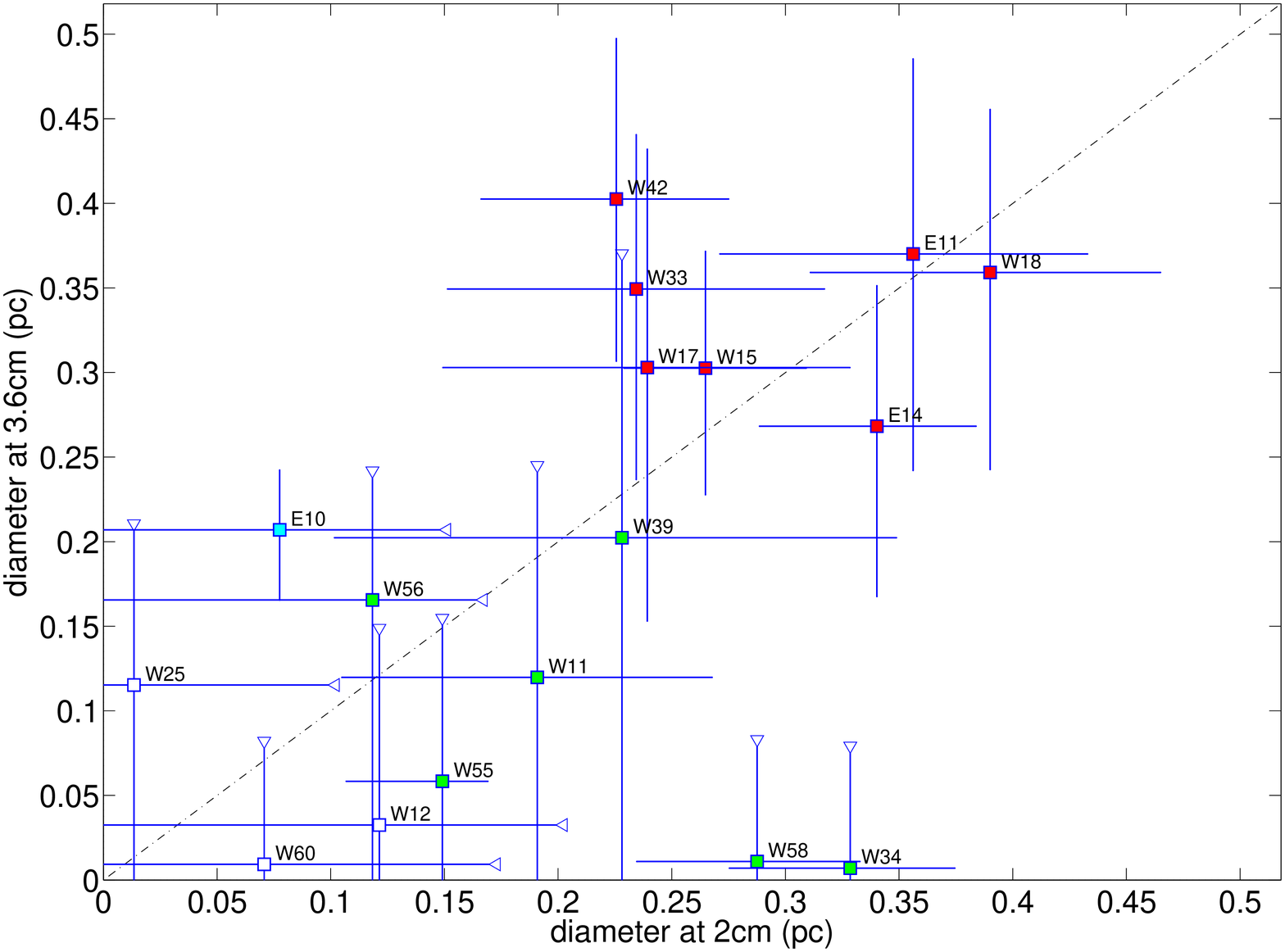}
\caption{Source diameter at 2~cm and 3.6~cm with error bars. Red sources are resolved at both bands, green only at 2~cm, blue only at 3.6~cm and white are unresolved. Red sources are classified as SNRs except E14.}
\label{fig:sizes} 
\end{center}
\end{figure}

\section{HSA 2~cm and global VLBI 3.6~cm observations}
\subsection{Source sizes and classes} \label{seq:sizes&classes}
High-resolution HSA 2~cm and global VLBI 3.6~cm observations of Arp220 performed in December 2006 resulted in  the detection of 17 sources in the Eastern and Western nuclei; a number of these sources are resolved - 13 at 2~cm, 8 at 3.6~cm, and 7 at both wavelengths (Batejat et al 2011 in preparation).
We find a good correlation between source sizes measured at 2~cm and 3.6~cm (Fig~\ref{fig:sizes}). Based on their time variability and spectra we classified each source as a SN or a SNR (see \S~\ref{sec:SNe&SNRs}). All SNRs detected at high frequencies are resolved with sizes $d > 0.27\ \mathrm{pc}$. Only one source classified as being a SN (E14) is resolved at both bands with a size $d = 0.30\ \mathrm{pc}$. All other sources have sizes $d < 0.22\ \mathrm{pc}$. Sizes are calculated combining the results of our observations at both bands. We conclude that our SNe have upper limits on their expansion velocities of $9\,000$ to $80\,000\ \mathrm{km~s^{-1}}$.

\subsection{Luminosity - Diameter relation} \label{sec:LDrel}
\begin{figure} [ht]
\begin{center}
\includegraphics[width=0.82\textwidth]{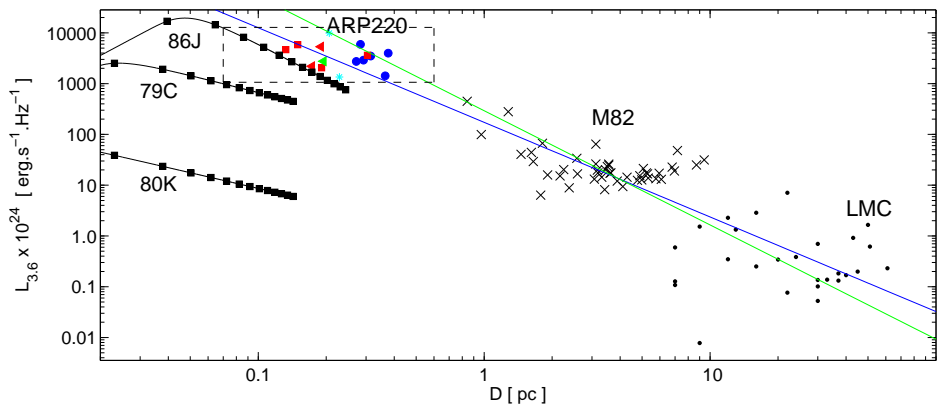}
\caption{Illustration of the 3.6~cm luminosities and sizes of observed SNe and SNRs. The SNe tracks for SN1986J (Type~IIn), SN1979C (Type~IIL) and SN1980K (Type~IIL) were produced using the light curve fits given by Weiler et al~\cite{Weiler02} combined with the deceleration parameter from Bietenholz et al~\cite{Bietenholz02} for SN1986J and assuming free expansion at $10^4\ \mathrm{km~s^{-1}}$ for both SN1980K and SN1979C. The square markers along each track indicate time evolution and are 1~year apart. Red points = sources identified as SNe, blue points~=~sources identified as SNRs, green point = possible transition object, cyan stars = unclassified sources.
The green line is the theoretical relation of Berezhko and V\"olk~\cite{BV04} of slope -9/4. The blue line is the best fit considering just those sources identified as SNRs. The data for the SNRs in M82 are taken from Huang et al~\cite{Huang94} and the data for the SNRs in the LMC from Mills et al~\cite{Mills84}.}
\label{fig:LDrelation} 
\end{center}
\end{figure}

Fig~\ref{fig:LDrelation} plots the detected Arp220 source sizes together with the SNRs detected in M82 and in the LMC. The best fit to our observations is a power law relationship between the luminosity and the diameter of the SNRs ($L \propto D^{-1.9}$)
close to the theoretical relation for SNRs in the Sedov phase as derived by Berezhko and V\"olk~\cite{BV04} ($L \propto D^{-9/4}$)
which also fits our data within the scatter (green line in Fig~\ref{fig:LDrelation}). This model is based on the assumption of an internal B-field energy being a constant small fraction of the post shock pressure ($\rho_{\mathrm{ISM}}.V_\mathrm{S}^2$) and therefore declining with the source size as the SNR expansion decelerates (see~\cite{BV04} for the details).
The data shown in Fig~\ref{fig:LDrelation} provides strong support for the existence of the Luminosity-Diameter relation with an exponent close to that theoretically predicted (Batejat et al 2011 in preparation).
Alternative models, such as the one presented by Thompson et al~\cite{Thompson09}, claim a constant internal B-field in the Sedov phase being simply proportional to the ISM \mbox{B-field} increased by a factor 3 to 6 and predict a constant radio luminosity versus size relation. This is inconsistent with our observed Luminosity-Diameter relation. Additionally in conflict to this theory the 18~cm flux of individual SNRs decline by $\sim 4\%/\mathrm{yr}$ (Batejat et al 2011 in preparation) which is consistent with Berezhko and V\"olk~\cite{BV04} model if the diameter of the sources expands by $2\%/\mathrm{yr}$ (as expected in Arp220 given their measured sizes and expected expansion velocities) confirming the idea of a B-field changing with source size.

\section{Recent global VLBI 6~cm monitoring}
We performed deep global VLBI 6~cm observations at three epochs on June 2008, October 2008 and February 2009 with the goal of looking for short time variability sources of all kinds, i.e. due to type Ib/c supernovae or AGN candidates, as well as producing the deepest 6~cm map to date of Arp220 by combining the three epochs with a noise rms of about $8\mu\mathrm{Jy/beam}$ (Fig~\ref{fig:exotic}).

\subsection{New detections} \label{sec:detections}
Nine sources are detected for the first time; five in the Eastern nucleus and four in the Western nucleus of Arp220. One of these (W61) seems to be a new rising source that was not present at the same flux density at earlier 6~cm epochs.

\subsection{Transition object candidate} \label{sec:transition}
Most detected objects are classified as SNe and SNRs. However this campaign of observations confirm the unusual characteristics of a few sources. One of them, referred as W12, has light curves which increase at all frequencies (3.6~cm, 6~cm and 18~cm) between 2004 and 2009. For SNRs we expect declining light-curves at all frequencies and for SNe light curves that rise at low frequencies and decline at high frequencies (see \S~\ref{sec:SNe&SNRs}). The observed behavior of W12 makes it a good candidate for being a SN/SNR transition object in which the shock front has just reached the ISM and electrons are being accelerated giving a rising flux density at all frequencies.

\subsection{Variable sources} \label{sec:variable}
\begin{figure} [ht]
\begin{center}
\includegraphics[width=0.65\textwidth]{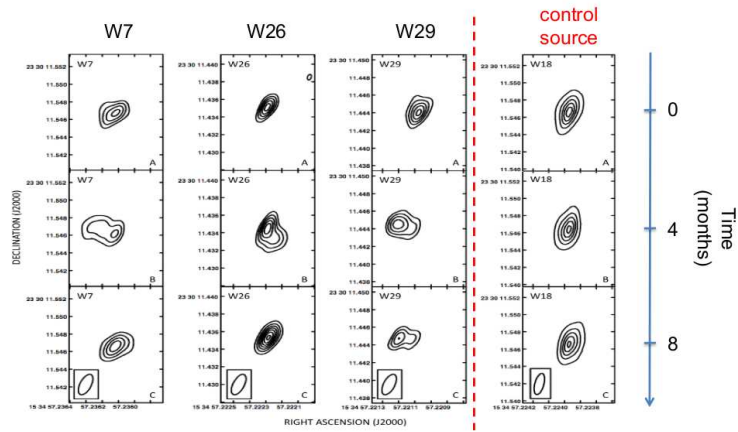}
\caption{Variability of W7, W26 and W29 over time at 5~GHz. The columns show in order the varying sources W7, W26 and W29 and finally as a control the non-varying resolved source W18 at epochs A to C (from top to bottom). The contours are at 4, 6, 8, ... times the rms noise for the variable sources and at 4, 12, 20,~... times the rms noise for W18. The rms noise is approximately $15\ \mu\mathrm{Jy/beam}$ at all 3 epochs. The beam is plotted for each source on the last row.}
\label{fig:variable} 
\end{center}
\end{figure}

We also have detected three rapidly variable sources. Those three sources, W7, W26 and W29, all localised in the Western nucleus of Arp220, are rapidly varying in flux at 6~cm and 18~cm, but also in position and shape with secondary features appearing and disappearing from epoch to epoch (Fig~\ref{fig:variable}). If emitted half way between epochs the speed estimates of the blob features are superluminal with speeds from $4c$ to $12c$, similar to the recently discovered source in M82 by Muxlow et al~\cite{Muxlow10}. The sources W7 and W26 are long-lived sources observed for more than 15~years. W29 is greater than 8 years but it is likely much older because it would not have been detected at the same flux density in earlier less sensitive epochs. All three sources show behaviours inconsistent with conventional SNe or SNRs.
\begin{figure} [ht]
\begin{center}
\includegraphics[width=0.66\textwidth]{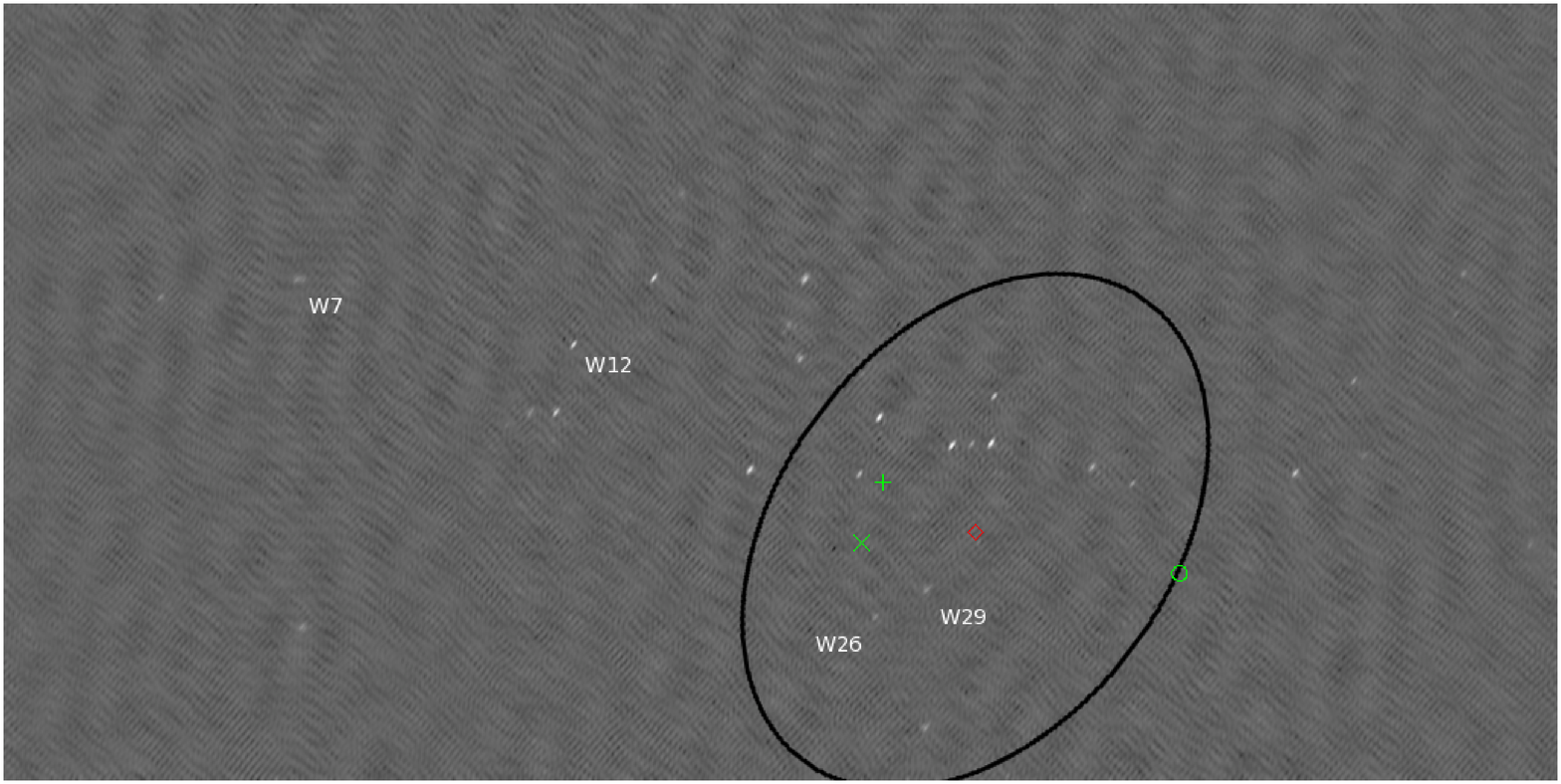}
\caption{Greyscale shows the combined 5~GHz global array GC031A,B,C natural weighted image of the Western nucleus. The rms noise is $8\mu\mathrm{Jy/beam}$, the image size is $512\times256\ \mathrm{mas}$. Labeled in white are the transition object candidate W12 and the highly variable sources W7, W26 and W29. The green crosses give the 1.3~mm and 2.3~mm IRAM positions of the hot dust "AGN" feature from Downes and Eckart~\cite{Downes07}, the green circle is the position from the 0.8~mm observation of Sakamoto et al~\cite{Sakamoto08}. Errors on these positions are estimated to be of order 50--100~mas. The black ellipse gives the orientation and size of the hot dust feature as fitted by Downes and Eckart~\cite{Downes07} centred at the centroid of the three mm interferometer positions.}
\label{fig:exotic} 
\end{center}
\end{figure}
It is possible that at least one of those sources is associated with a supermassive black hole (SMBH). Downes and Eckart~\cite{Downes07} presented dynamical evidence for such a SMBH in Arp220's Western nucleus by analysing the position-velocity diagram from CO(2-1) observations. Downes and Eckart~\cite{Downes07} and Sakamoto et al~\cite{Sakamoto08} have analysed the mm continuum emission from hot dust and have argued that it may be too compact to be from a starburst and could be black hole powered instead.
When comparing the position and size of the hot-dust feature from Downes and Eckart~\cite{Downes07} and Sakamoto et al~\cite{Sakamoto08} with the positions of our three variable objects (see Fig~\ref{fig:exotic}) we see that W26 and W29 are located close to the centre of the hot dust emission making them possible AGN candidates.
It is however difficult to come to a final conclusion on the true nature of those three variable sources.
Are those objects accreting supermassive black holes? Could they be highly beamed stellar-mass black holes
(`micro-blazars') or even radio-loud versions of accreting intermediate-mass black holes~\cite{Farrell09}.
Scheduled HSA observations in the spring 2011 looking for variability on timescales shorter than 30~days will hopefully give us more clues as to the nature of these objects.

\section{Conclusion}
The two observation campaigns presented in this work allow us to conclude that the compact radio sources detected in Arp220 are a mixture of SNe and SNRs. We also confirm the theoretical Berezhko and V\"olk~\cite{BV04} Luminosity-Diameter relation for SNRs and extend it to very young sources. We have possibly detected a SN/SNR transition object and we have detected at least three highly variable sources whose structures vary on timescales shorter than 4~months with possible superluminal motion.

\acknowledgments
The European VLBI Network is a joint facility of European, Chinese, South African and other radio astronomy institutes funded by their national research councils.
The National Radio Astronomy Observatory is a facility of the National Science Foundation operated under cooperative agreement by Associated Universities, Inc.

\end{document}